\documentclass[twocolumn,showpacs,preprintnumbers,amsmath,amssymb]{revtex4}
\usepackage{graphicx}% Include figure files
\usepackage{dcolumn}% Align table columns on decimal point
\usepackage{bm}% bold math
\begin{document}

%\preprint{}

\title{Strongly interacting and highly entangled photons in asymmetric quantum well with resonant tunneling}
% Force line breaks with \\

\author{Shuangli Fan$^{1}$}
\author{Hui Sun$^{1,2,4}$}\thanks{physunh@snnu.edu.cn}
\author{Xun-Li Feng$^{2}$}
\author{Chunfeng Wu$^{2}$}
\author{Shangqing Gong$^{3}$}
\author{Guoxiang Huang$^{4}$}
\author{C. H. Oh$^{2}$}\thanks{phyohch@nus.edu.sg}

\affiliation{$^{1}$ School of Physics and Information Technology,
Shaanxi Normal University, Xi'an 710062, People's Republic of
China\\
$^{2}$ Centre for Quantum Technologies and Department of Physics,
National University of Singapore, 3 Science Drive 2, Singapore 117543\\
$^{3}$ Department of Physics, East China University of Science and
Technology, Shanghai 200237, People¡¯s Republic of China\\
$^{4}$ State Key Laboratory of Precision Spectroscopy, East China
Normal University, Shanghai 200062, China}

\date{\today}% It is always \today, today,

\begin{abstract}
We propose an asymmetric quantum well structure to realize strong
interaction between two slow optical pulses. The linear optical
properties and nonlinear optical responses associated with
cross-Kerr nonlinearity are analyzed. Combining the resonant
tunneling and the advantages of inverted-Y type scheme, giant
cross-Kerr nonlinearity can be achieved with vanishing absorptions.
Based on the unique feature, we demonstrate that highly entangled
photons can be produced and photonic controlled phase gate can be
constructed. In this construction, the scheme is symmetric for the
probe and signal pulses. Consequently, the condition of group
velocity matching can be fulfilled by adjusting the initial electron
distribution.
\end{abstract}

\pacs{78.67.De, 42.65.Hw}% PACS, the Physics and Astronomy
%\keywords{ laser beam, plasma channel, NPA, HOR}%Use showkeys class option if keyword
\maketitle

\section{introduction}

Photons are ideal carriers of quantum information as they do not
interact strongly with their environment and can be transmitted over
long distances~\cite{nielsen-book,kok-rmp-2007}. Realizing efficient
nonlinear interactions between single photons is considered a key
step toward all-optical quantum computation and quantum information
processing. While nonlinear effect whereby one light beam influences
another requires large numbers of photons or else photon will be
confined in a high-$Q$ cavity. Hence the major obstacle of
constructing scalable and efficient quantum computation with
photonic qubits is the absence of giant cross-Kerr nonlinearity
capable of entangling pairs of photons. A promising avenue has been
opened by studies of enhanced nonlinear coupling via
electromagnetically induced transparency
(EIT)~\cite{harris-prl-1999,fleischhauer-rmp-2005}. In a four-level
$N$-type scheme, it was proposed that the ultrahigh sensitivity of
EIT dispersion to the two-photon Raman detuning in the vicinity of
an absorption minimum can be used to enhance cross-Kerr nonlinearity
between two weak optical
fields~\cite{schmidt-ol-1996,fleischhauer-rmp-2005}. Large
cross-Kerr nonlinearity emerges when two optical pulses, a probe and
a signal, interact for a sufficiently long time. This happens when
their group velocities are both reduced and
comparable~\cite{lukin-prl-2000,ottaviania-ejpd-2006,ottaviani-prl-2003}.
In order to eliminate the mismatch between the slow group velocity
of the probe pulse subject to EIT and that of the nearly free
propagating signal pulse, versatile novel symmetric configurations
have been suggested theoretically and experimentally to realize the
polarization phase gate, including the tripod
configuration~\cite{rebic-pra-2004,petrosyan-pra-2004,petrosyan-job-2005,
li-prl-2008,guo-jpb-2011}, inverted-Y-type
configuration~\cite{joshi-pra-2005,bai-oc-2010}, M-type atomic
schemes~\cite{ottaviani-prl-2003,ottaviania-ejpd-2006,shiau-prl-2011,li-pla-2006},
and so on~\cite{hou-pra-2009}. More recently, large and rapidly
responding cross-Kerr nonlinearity and highly entangled photons have
been demonstrated in resonant atomic scheme based on active Raman
gain configurations~\cite{deng-prl-2007,hang-pra-2010,hang-oe-2010}.

Semiconductor heterostructures provide a potential energy well with
a size comparable to the de Broglie wavelength, trapping the
carriers in discrete energy levels resulting in objects with
atom-like optical properties. Different from atomic system, the
interaction between semiconductor heterostructures and optical
fields is strongly enhanced with merits such as the large electric
dipole moments due to the small effective electron mass. Moreover,
the intersubband energies and the electron function symmetries can
be engineered as desired in accordance with the requirement. These
advantages create the opportunities of building opto-electron
devices that harness atom physics. Another important motivation of
such study comes from the drastic increase in applications because
of the wide-spread use of semiconductor components in
optoelectronics and quantum information science. As a consequence,
there has been a fast growth of research activity aimed at studying
the quantum interference effects in semiconductors, for examples,
the strong EIT~\cite{phillips-prl-2003}, tunneling induced
transparency (TIT)~\cite{schmidt-apl-1997}, ultrafast optical
switching with Fano interference~\cite{wu-prl-2005}, slow
light~\cite{yuan-apl-2006},
etc~\cite{paspalakis-prb-2004,paspalakis-prb-2006-1,paspalakis-prb-2006-2,wu-pra-2006,sun-prb-2009-2}.
Nonlinear optical properties in semiconductor heterostructures have
also been paid much attention such as ultraslow optical solitons
with TIT~\cite{zhu-prb-2009,yang-pra-2008}, enhancement of self-Kerr
nonlinearity~\cite{kosionis-jap-2011,sun-prb-2006}, controlled phase
shift up to $\pi/4$ in a single-quantum dot coupled to a photonic
crystal nanocavity~\cite{fushman-science-2008}, giant cross-Kerr
nonlinearity with spin-orbit coupling~\cite{sun-prb-2009-1}, and so
on~\cite{li-prb-2011}. Recently, the realization of giant cross-Kerr
nonlinear phase shift and the related quantum information processing
(QIP) has been investigated in quantum well (QW) structures based on
interband and intersubband
transitions~\cite{yang-oe-2008,hao-josab-2010}.

In QW structure, resonant tunneling can induce not only transparency
but also large cross-Kerr nonlinearity~\cite{sun-ol-2007}. However,
the nonlinear phase shift on order of $\pi$ cannot be achieved at
single-photon level, since the group velocities of the probe and
signal pulses are mismatched~\cite{sun-ol-2007}. This fact limits
its applications in QIP. In the present paper, we suggest an
alternative asymmetric QW structure, which combines resonant
tunneling and the advantages of inverted-Y-type configuration, and
study the linear optical properties and nonlinear optical responses
associated with cross-Kerr nonlinearity. It is found, in the present
QW structure, that giant cross-Kerr nonlinearity can be achieved
with vanishing linear and nonlinear absorptions simultaneously.
Consequently, highly entangled photons can be produced and
polarization photonic controlled phase gate can be constructed. More
importantly, for the probe and signal pulses, the structure is an
inherent symmetric configuration. Hence the condition of group
velocity matching can be easily satisfied by adjusting the initial
electron distribution.

\begin{figure}
\includegraphics[width=7cm]{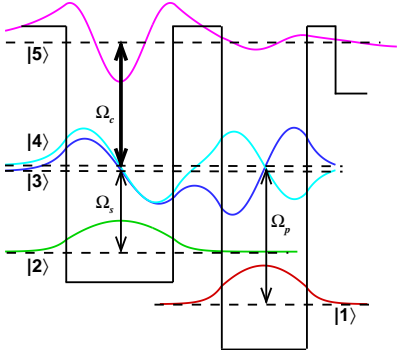}
\caption{(color online) Conduction subband of the asymmetric quantum
well structure. The solid curves represent the corresponding wave
functions.}\label{fig-band-structure}
\end{figure}

\section{structure and linear optical properties}

Our asymmetric double QW structure is shown in
Fig.~\ref{fig-band-structure}. The growth sequence of the structure
from left to right is as follows. A thick Al$_{0.50}$Ga$_{0.50}$As
barrier is followed by an Al$_{0.10}$Ga$_{0.90}$As layer with
thickness of 8.8~nm (shallow well). This shallow well is separated
from a 6.9~nm GaAs layer (deep well) on the right by a 3.8~nm
Al$_{0.50}$Ga$_{0.50}$As potential barrier. Finally, a thin (2.4~nm)
Al$_{0.50}$Ga$_{0.50}$As barrier separates the deep well from the
last Al$_{0.40}$Ga$_{0.60}$As thick layer of on the right. In this
structure, one would observe the ground subbands of the right deep
well $|1\rangle$ and the left shallow well $|2\rangle$ with energies
57.2~meV and 123.1~meV, respectively. The eigenenergy of the second
excited subband of the left shallow well $|5\rangle$ is 385.9~meV.
Two new subbands $|3\rangle$ and $|4\rangle$ with eigenenergies
$224.1$~meV and $231.4$~meV are, respectively, created by mixing the
first excited subbands of the shallow ($|se\rangle$) and deep
($|de\rangle$) wells by tunneling. Their corresponding wave
functions are symmetric and antisymmetric combinations of
$|se\rangle$ and $|de\rangle$, i.e.,
$|3\rangle=(|se\rangle-|de\rangle)/\sqrt{2}$ and
$|4\rangle=(|se\rangle+|de\rangle)/\sqrt{2}$. The basic idea is to
combine resonant tunneling with the inherited symmetry of
invert-Y-type configuration. To do so, we apply a weak probe and a
weak signal fields with frequencies $\omega_{p}$ and $\omega_{s}$ to
drive the transitions $|1\rangle\leftrightarrow|3\rangle$,
$|1\rangle\leftrightarrow|4\rangle$ and
$|2\rangle\leftrightarrow|3\rangle$,
$|2\rangle\leftrightarrow4\rangle$, respectively. The subbands
$|3\rangle$ and $|4\rangle$ are coupled with $|5\rangle$ by a
continuous-wave control field with angular frequency $\omega_{c}$.
Thus, an inverted-Y-type configuration with two-fold degenerate
middle subbands is realized. Under the dipole and rotating-wave
approximations (RWA), this structure is governed by a set of density
matrix equations given below,
\begin{eqnarray}
&&\hspace{-0.4cm}\dot{\sigma}_{21}=id_{21}\sigma_{21}-i\Omega_{p}(\sigma_{23}+m\sigma_{24})
+i\Omega_{s}(\sigma_{31}+q\sigma_{41}),\label{density-eq-21}\\
&&\hspace{-0.4cm}\dot{\sigma}_{31}=id_{31}\sigma_{31}+i\Omega_{p}(\sigma_{11}-\sigma_{33})-im\Omega_{p}\sigma_{34}
+i\Omega_{s}\sigma_{21}\nonumber\\
&&\hspace{0.6cm}+i\Omega_{c}\sigma_{51},\label{density-eq-31}\\
&&\hspace{-0.4cm}\dot{\sigma}_{41}=id_{41}\sigma_{41}+im\Omega_{p}(\sigma_{11}-\sigma_{44})-i\Omega_{p}\sigma_{43}
+iq\Omega_{s}\sigma_{21}\nonumber\\
&&\hspace{0.6cm}+ik\Omega_{c}\sigma_{51},\label{density-eq-41}\\
&&\hspace{-0.4cm}\dot{\sigma}_{51}=id_{51}\sigma_{51}-i\Omega_{p}(\sigma_{53}+m\sigma_{54})
+i\Omega_{c}(\sigma_{31}+k\sigma_{41}),\label{density-eq-51}\\
&&\hspace{-0.4cm}\dot{\sigma}_{32}=id_{32}\sigma_{32}+i\Omega_{s}(\sigma_{22}-\sigma_{33})+i\Omega_{p}\sigma_{12}
-iq\Omega_{s}\sigma_{34}\nonumber\\
&&\hspace{0.6cm}+i\Omega_{c}\sigma_{52},\label{density-eq-32}\\
&&\hspace{-0.4cm}\dot{\sigma}_{42}=id_{42}\sigma_{42}+iq\Omega_{s}(\sigma_{22}-\sigma_{44})+im\Omega_{p}\sigma_{12}
-i\Omega_{s}\sigma_{43}\nonumber\\
&&\hspace{0.6cm}+ik\Omega_{c}\sigma_{52},\label{density-eq-42}\\
&&\hspace{-0.4cm}\dot{\sigma}_{52}=id_{52}\sigma_{52}-i\Omega_{s}(\sigma_{53}+q\sigma_{54})
+i\Omega_{c}(\sigma_{32}+k\sigma_{42}),\label{density-eq-52}
\end{eqnarray}
where $d_{21}=\Delta_{p}-\Delta_{s}+i\gamma_{21}$,
$d_{31}=\Delta_{p}+i\gamma_{31}$,
$d_{41}=\Delta_{p}-\delta+i\gamma_{41}$,
$d_{51}=\Delta_{p}+\Delta_{c}-\Delta_{s}+i\gamma_{51}$,
$d_{32}=\Delta_{s}+i\gamma_{32}$,
$d_{42}=\Delta_{s}-\delta+i\gamma_{42}$,
$d_{52}=\Delta_{c}+i\gamma_{52}$ with $\Delta_{p}$, $\Delta_{s}$,
and $\Delta_{c}$ being the detunings of the probe, signal and
control fields with the corresponding transitions, and they are
defined as
$\Delta_{p,(s,c)}=\omega_{p,(s,c)}-(\omega_{3,(3,5)}-\omega_{1,(2,3)})$.
$\delta=\omega_{4}-\omega_{3}\simeq7.3$~meV denotes the energy
difference between the subbands $|3\rangle$ and $|4\rangle$. Halves
of the Rabi frequencies of the probe, signal and control fields are
$\Omega_{p}=\vec{\mu}_{13}\cdot\vec{E}_{p}/2\hbar$,
$\Omega_{s}=\vec{\mu}_{23}\cdot\vec{E}_{s}/2\hbar$, and
$\Omega_{c}=\vec{\mu}_{53}\cdot\vec{E}_{c}/2\hbar$ with
$\vec{\mu}_{ij}$ being electric dipole momentum between subbands
$|i\rangle$ and $|j\rangle$ ($i,j=1-5$ and $i\neq j$), while
$m=\mu_{41}/\mu_{31}=-0.73$, $q=\mu_{42}/\mu_{32}=1.2$, and
$k=\mu_{54}/\mu_{53}=2.3$ give the ratios between the relevant
subband transition dipole momentum. $E_{p}$, $E_{s}$, and $E_{c}$
are, respectively, the slowly varying electric field amplitudes of
the probe, signal and control fields. The half linewidths are,
respectively, given by $\gamma_{31}=\gamma_{3}+\gamma_{31}^{\rm
deph}$, $\gamma_{41}=\gamma_{4}+\gamma_{41}^{\rm deph}$,
$\gamma_{51}=\gamma_{5}+\gamma_{51}^{\rm deph}$,
$\gamma_{32}=\gamma_{3}+\gamma_{32}^{\rm deph}$,
$\gamma_{42}=\gamma_{4}+\gamma_{42}^{\rm deph}$,
$\gamma_{52}=\gamma_{5}+\gamma_{52}^{\rm deph}$. Here $\gamma_{3}$
($\gamma_{4}$, $\gamma_{5}$) is the electron decay rate of subband
$|3\rangle$, ($|4\rangle$, $|5\rangle$) and $\gamma_{ij}^{\rm deph}$
the electron dephasing rates, which are introduced to account not
only for intrasubband phonon scattering and electron-electron
scattering but also for inhomogeneous broadening due to scattering
on interface roughness. The dipole transition rate from subband
$|2\rangle$ to $|1\rangle$ is very small because of the high
inter-well barrier between them, $\gamma_{21}\approx\gamma_{21}^{\rm
deph}$. Electron decay rates can be calculated by solving effective
mass Schr\"{o}dinger equation~\cite{ahn-prb-1986}. For temperature
up to 10~K and electron density smaller than $10^{12}$~cm$^{-2}$,
$\gamma_{i}^{\rm deph}$ can be estimated according to
Ref.~\cite{schmidt-apl-1997}.

\begin{figure}
\includegraphics[width=8cm]{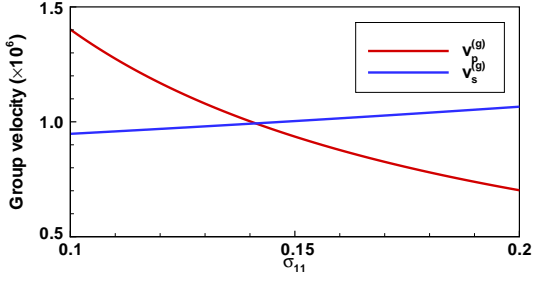}
\caption{(color online) The group velocities of the probe and the
signal pulses as functions of $\sigma_{11}^{(0)}$ with and without
the control field. The parameters are explained in the
text.}\label{fig-group-velocity}
\end{figure}

If all electrons remain in ground subband $|1\rangle$, which means
that the signal field drives two virtually empty transitions, the
contribution to the susceptibility comes only from higher order. It
is hard to achieve group velocity matching since the asymmetry of
configuration. Thus, as done in Ref.~\cite{sun-prb-2009-1}, we
assume that electrons distribute not only in subbands $|1\rangle$
but also in $|2\rangle$. The symmetric configuration is hence
formed. In the presence of the control field, the subbands
$|3,4\rangle$ and $|5\rangle$ are mixed into three new subbands. The
symmetry of scheme ensures that the probe and signal propagate with
comparable group velocity. To investigate the group velocities of
the probe and signal pulses, following the standard
processes~\cite{hang-oe-2010}, we assume
$|\Omega_{p}|,|\Omega_{s}|\ll|\Omega_{c}|,\Delta_{1},\Delta_{2},\Delta_{3},\delta$
and solve the density matrix equations
(\ref{density-eq-21})-(\ref{density-eq-52}) in the nondepletion
approximation ($\sigma_{11}+\sigma_{22}\approx1$) together with
Maxwell's equations and expand the linear dispersion relations as
Taylor series around their center frequency. The group velocities of
the probe and signal pulses are, respectively, given by
\begin{eqnarray}
&&v_{p}^{g}=1/{\rm Re}[K_{p}^{(1)}], \hspace{0.6cm}v_{s}^{g}=1/{\rm
Re}[K_{s}^{(1)}],\label{eq-vg}
\end{eqnarray}
with \begin{widetext}\begin{eqnarray}
&&\hspace{-0.5cm}K_{p}^{(1)}=\frac{1}{c}+\frac{N\omega_{p}|\mu_{31}|^{2}\sigma_{11}^{(0)}}{\hbar\epsilon_{0}c}
\left\{\frac{[d_{31}d_{41}+d_{31}d_{51}+d_{41}d_{51}-(1+k^{2})\Omega_{c}^{2}]
[d_{51}(d_{41}+m^{2}d_{31})-(k-m)^{2}\Omega_{c}^{2}]}
{[d_{31}d_{41}d_{51}-(d_{41}+k^{2}d_{31})\Omega_{c}^{2}]^{2}}\right.\nonumber\\
&&\left.\hspace{4cm}-\frac{d_{41}+m^{2}d_{31}+(1+m^{2})d_{51}}{d_{31}d_{41}
d_{51}-(d_{41}+k^{2}d_{31})\Omega_{c}^{2}}\right\},\label{eq-vg-p}\\
&&\hspace{-0.5cm}K_{s}^{(1)}=\frac{1}{c}+\frac{N\omega_{s}|\mu_{32}|^{2}\sigma_{22}^{(0)}}{\hbar\epsilon_{0}c}
\left\{\frac{[d_{32}d_{42}+d_{32}d_{52}+d_{42}d_{52}-(1+k^{2})\Omega_{c}^{2}]
[d_{52}(d_{42}+m^{2}d_{32})-(k-q)^{2}\Omega_{c}^{2}]}
{[d_{32}d_{42}d_{52}-(d_{42}+k^{2}d_{32})\Omega_{c}^{2}]^{2}}\right.\nonumber\\
&&\left.\hspace{4cm}-\frac{d_{42}+m^{2}d_{32}+(1+m^{2})d_{52}}{d_{32}d_{42}
d_{52}-(d_{42}+k^{2}d_{32})\Omega_{c}^{2}}\right\},\label{eq-vg-s}
\end{eqnarray}\end{widetext}
where $N$ is the electron volume density, $\sigma_{11}^{(0)}$ and
$\sigma_{22}^{(0)}$ are initial electron distribution in subbands
$|1\rangle$ and $|2\rangle$ with
$\sigma_{11}^{(0)}+\sigma_{22}^{(0)}=1$. The dependence of
$v_{p}^{g}$ and $v_{s}^{g}$ on $\sigma_{11}^{(0)}$ shows that it is,
in principle, possible to control the group velocities by adjusting
the initial electron distribution. The electron decay rates are
$\gamma_{3}\approx\gamma_{4}=0.5$~meV, $\gamma_{5}=0.2$~meV
(corresponding intrasubband relaxation time
$T_{1}\sim10$~ps)~\cite{williams-prl-2001} and $\gamma_{3}^{\rm
deph}=\gamma_{4}^{\rm deph}=\gamma_{5}^{\rm
deph}=0.2$~meV~\cite{schmidt-apl-1997}. We take the Rabi frequency
and the detuning of the control field as $\Omega_{c}=1.5$~meV and
$\Delta_{c}=-5.3$~meV. With $\Delta_{p}=\Delta_{s}=3.0$~meV (around
the center of their transparency windows) and
$N=5\times10^{-17}$~cm$^{-3}$, Fig.~\ref{fig-group-velocity}
illustrates the dependence of the group velocities of the probe and
signal pulses on the initial electron distribution
$\sigma_{11}^{(0)}$. By controlling the initial electron
distribution ($\sigma_{11}^{(0)}\approx0.141$), the probe and signal
pulses will propagate with comparable and small group velocities
($v_{p}^{g}=v_{s}^{g}\approx1.0\times10^{6}$~m/s). The initial
electron distribution can be realized with stimulated Raman
adiabatic passage~\cite{jin-pra-2004}.

What we pursue is to produce the strongly interacting and highly
entangled photons by virtue of the giant cross-Kerr nonlinearity of
the QW structure considered, we hence define the susceptibility
as\cite{ottaviania-ejpd-2006}
\begin{eqnarray}
&&\chi_{p}=\frac{N|\mu_{13}|^{2}}{\hbar\epsilon_{0}}\frac{\sigma_{31}+m\sigma_{41}}{\Omega_{p}}\nonumber\\
&&\hspace{0.5cm}\simeq\chi_{p}^{(1)}+\chi_{p}^{\rm(3,
SPM)}|E_{p}|^{2} +\chi_{p}^{\rm
(3,XPM)}|E_{s}|^{2},\label{eq-chi-definition-p}
\end{eqnarray}
\begin{eqnarray}
&&\chi_{s}=\frac{N|\mu_{23}|^{2}}{\hbar\epsilon_{0}}\frac{\sigma_{32}+q\sigma_{42}}{\Omega_{s}}\nonumber\\
&&\hspace{0.5cm}\simeq\chi_{s}^{(1)}+\chi_{s}^{\rm(3,SPM)}|E_{s}|^{2}
+\chi_{s}^{\rm(3,XPM)}|E_{p}|^{2},\label{eq-chi-definition-s}
\end{eqnarray}
where $\chi_{p,s}^{(1)}$, $\chi_{p,s}^{\rm(3,SPM)}$ and
$\chi_{p,s}^{\rm(3,XPM)}$ are the linear, self-Kerr, and cross-Kerr
susceptibilities of the probe and signal pulses, respectively. By
solving the set of density matrix
equations~(\ref{density-eq-21})-(\ref{density-eq-52}) in steady
state in the nondepletion approximation, the first and third order
susceptibilities associated with cross-Kerr nonlinearity can be
calculated ($\chi_{p,s}^{\rm(3,XPM)}$ will be considered in the next
section). The linear susceptibilities can be written as
\begin{eqnarray}
\chi_{p}^{(1)}=\frac{N|\mu_{31}|^{2}}{\hbar\epsilon_{0}}
\chi_{p}^{'(1)},\hspace{0.3cm}
\chi_{s}^{(1)}=\frac{N|\mu_{32}|^{2}}{\hbar\epsilon_{0}}
\chi_{s}^{'(1)},
\end{eqnarray}
in which $\chi_{p}^{'(1)}$ and $\chi_{s}^{'(1)}$ are given by
\begin{eqnarray}
&&\chi_{p}^{'(1)}=-\sigma_{11}^{(0)}
\frac{d_{51}(d_{41}+m^{2}d_{31})-(k-m)^{2}\Omega_{c}^{2}}{d_{31}d_{41}d_{51}
-(d_{41}+k^{2}d_{31})\Omega_{c}^{2}},\label{eq-chi1-p}\\
&&\chi_{s}^{'(1)}=-\sigma_{22}^{(0)}
\frac{d_{52}(d_{42}+q^{2}d_{32})-(k-q)^{2}\Omega_{c}^{2}}{d_{32}d_{42}d_{52}
-(d_{42}+k^{2}d_{32})\Omega_{c}^{2}}.\label{eq-chi1-s}
\end{eqnarray}
Equations (\ref{eq-chi1-p})-(\ref{eq-chi1-s}) show the symmetry of
the QW structure between the probe and signal fields. With the
simultaneous exchange of $1\leftrightarrow2$ and $m\leftrightarrow
q$, the expression of $\chi_{s}^{(1)}$ can be obtained.

\begin{figure}
\includegraphics[width=7cm]{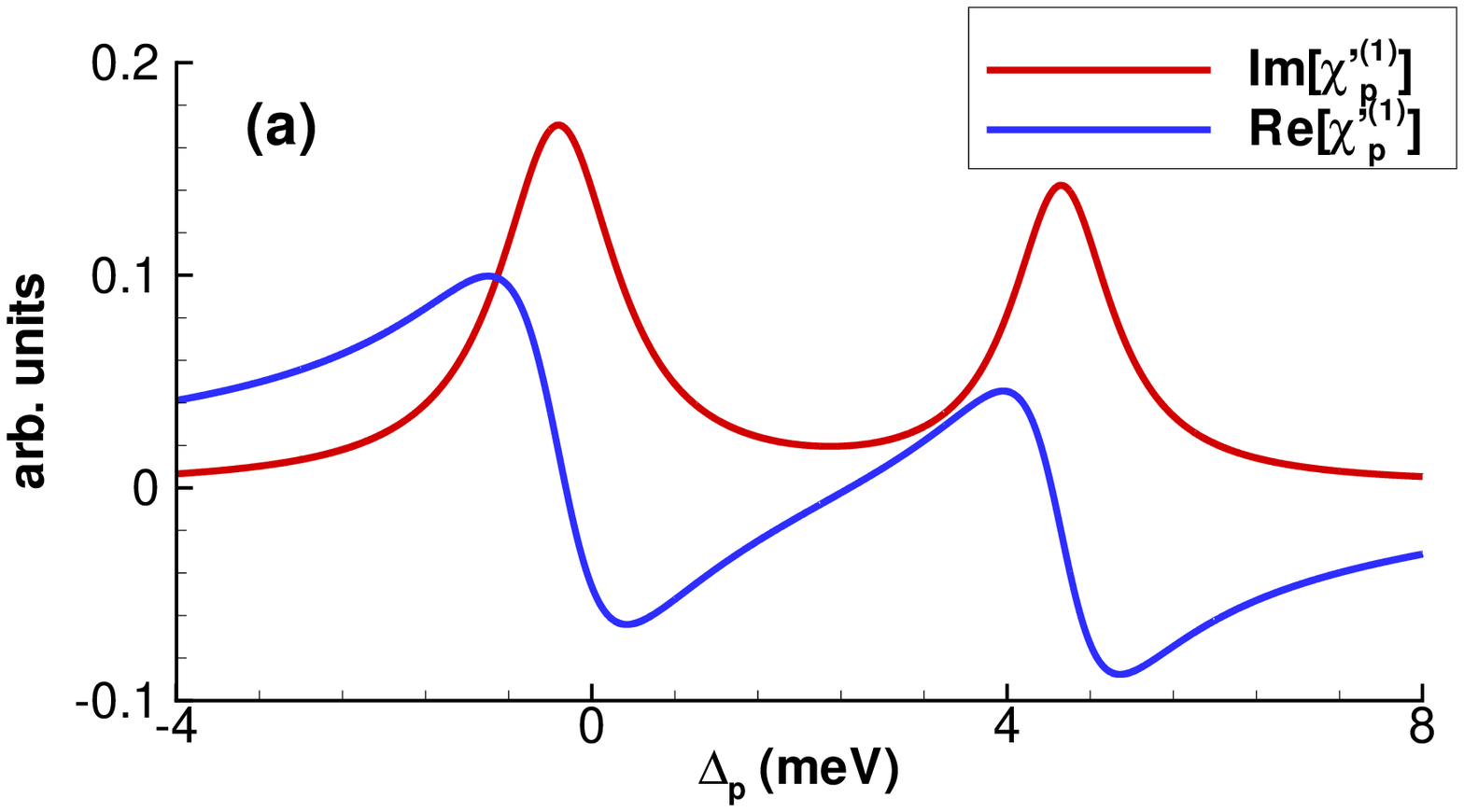}
\includegraphics[width=7cm]{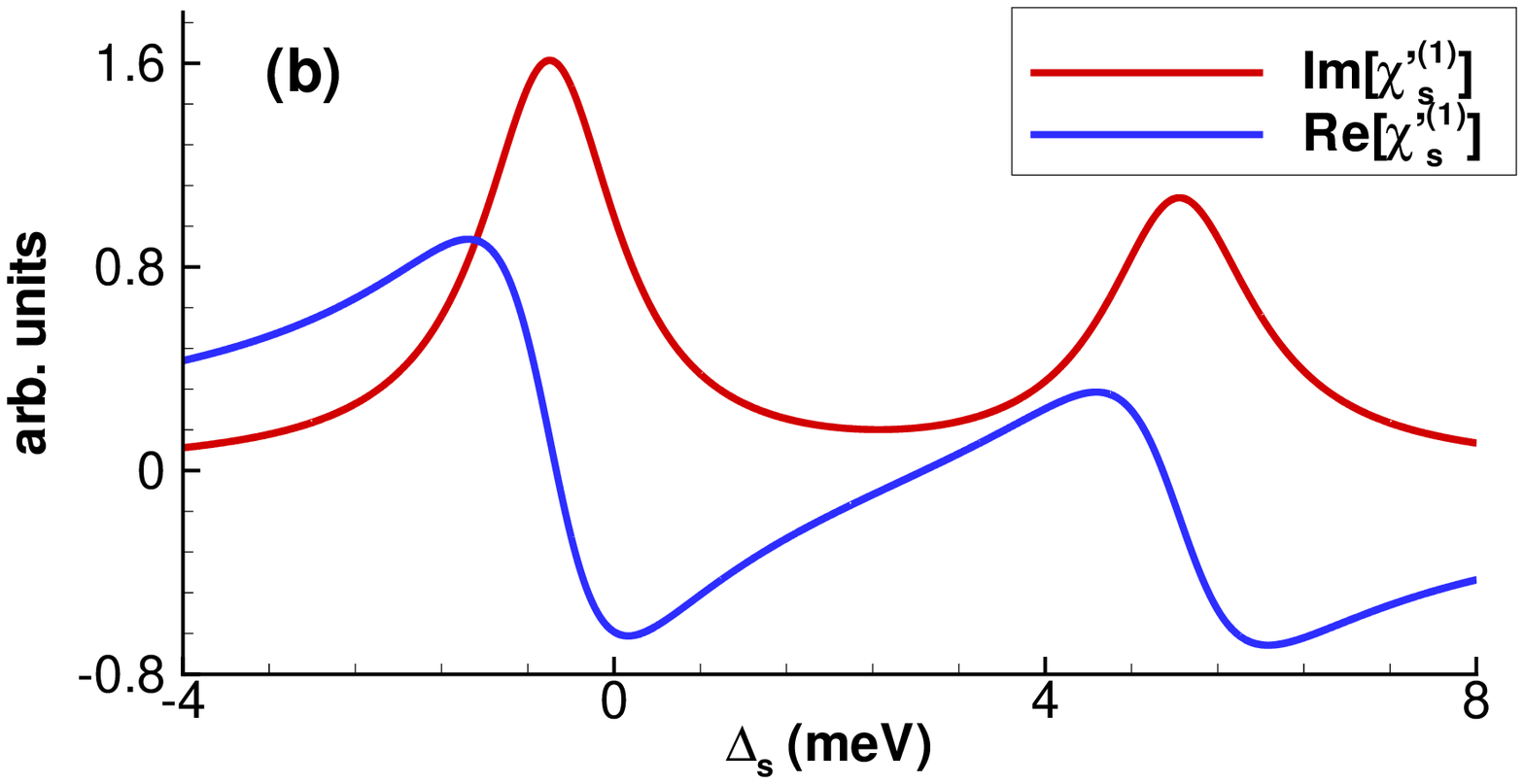}
\caption{(color online) The linear absorption (red curve) and
dispersion (blue curve) of the probe (a) and the signal fields (b)
as functions of their corresponding detunings $\Delta_{p}$ and
$\Delta_{s}$ with $\sigma_{11}^{(0)}\approx0.27$. The detunings are
chosen as (a) $\Delta_{s}=\delta/2=3.65$~meV, (b)
$\Delta_{p}=3.6505$~meV. The other parameters are the same with
those in
Fig.~\ref{fig-group-velocity}.}\label{fig-linear-properties}
\end{figure}

The real and imaginary parts of $\chi_{p}^{'(1)}$
($\chi_{s}^{'(1)}$), respectively, account for the linear absorption
and dispersion of the probe (signal) field. With
$\sigma_{11}^{(0)}\approx0.141$, their evolutions versus their
corresponding detunings are shown in
Figs.~\ref{fig-linear-properties}(a) and (b), respectively. We take
$\Delta_{s}=3.0$~meV in Fig.~\ref{fig-linear-properties}(a) and
$\Delta_{p}=2.995$~meV in Fig.~\ref{fig-linear-properties}(b). The
other parameters are the same with those in
Fig.~\ref{fig-group-velocity}. With this set of parameters, the
dispersion of the probe and the signal pulses around the center of
the transparent window are linearly proportional to their detunings.
This means that the probe and signal pulses will propagate with
comparable and small group velocities, and the influence of group
velocity dispersion can be neglected within the region considered.
At the center of the transparency window, the linear absorptions of
the probe and signal fields are very small because of the
destructive interference between transition pathes, and can be
safely ignored.

\section{strongly interacting and highly entangled photons with resonant tunneling}

\begin{figure*}
\includegraphics[width=7cm]{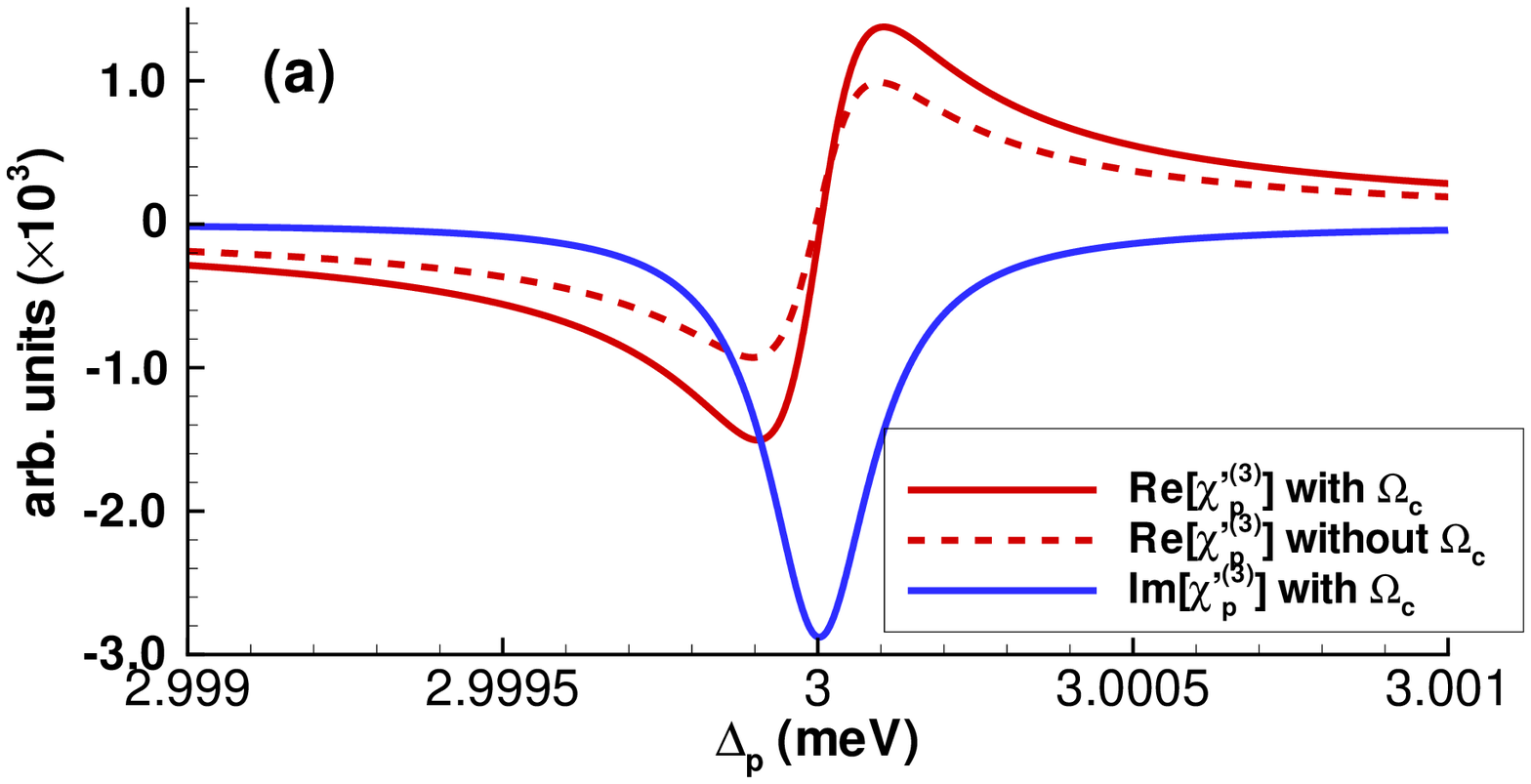}
\includegraphics[width=7cm]{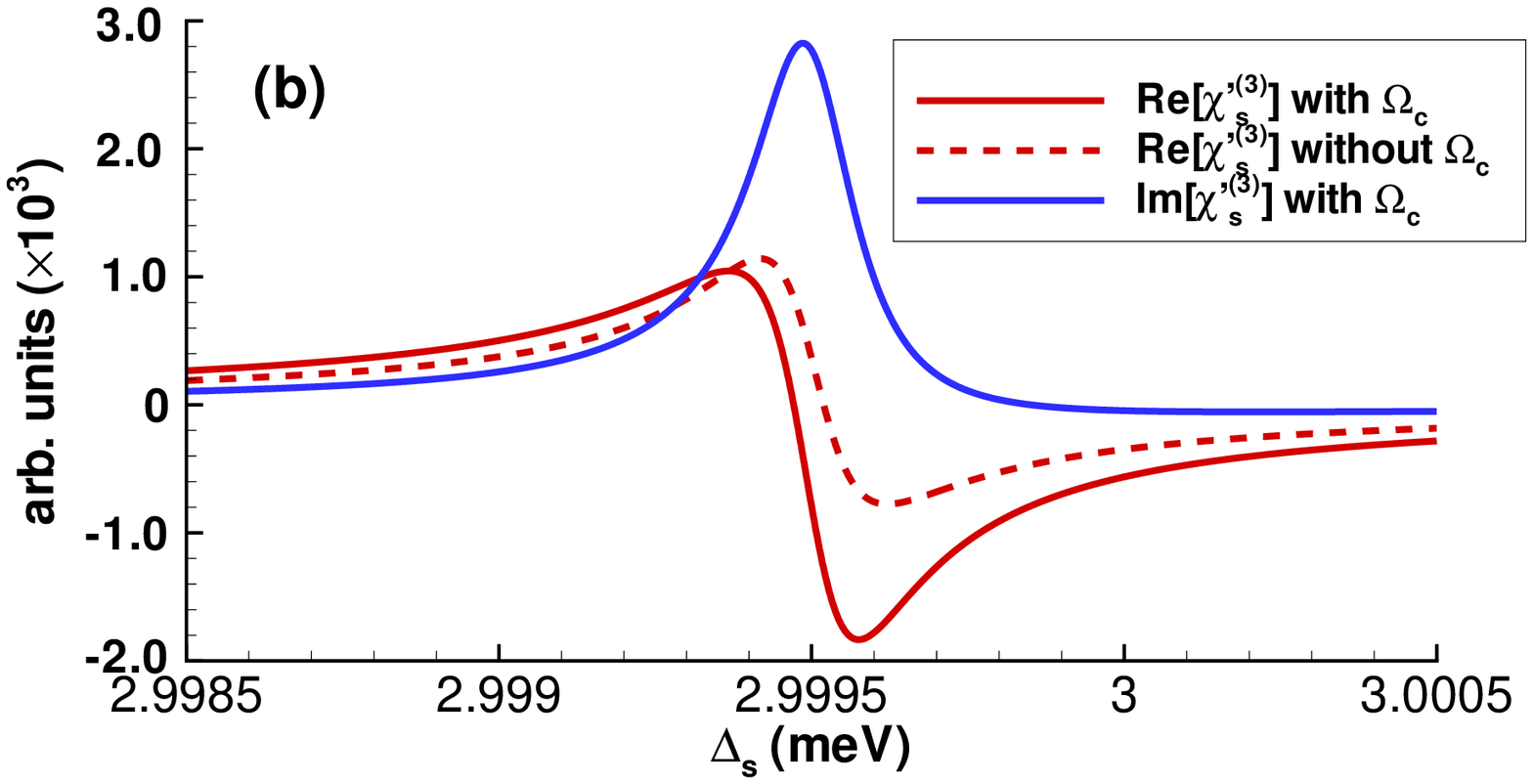}
\includegraphics[width=7cm]{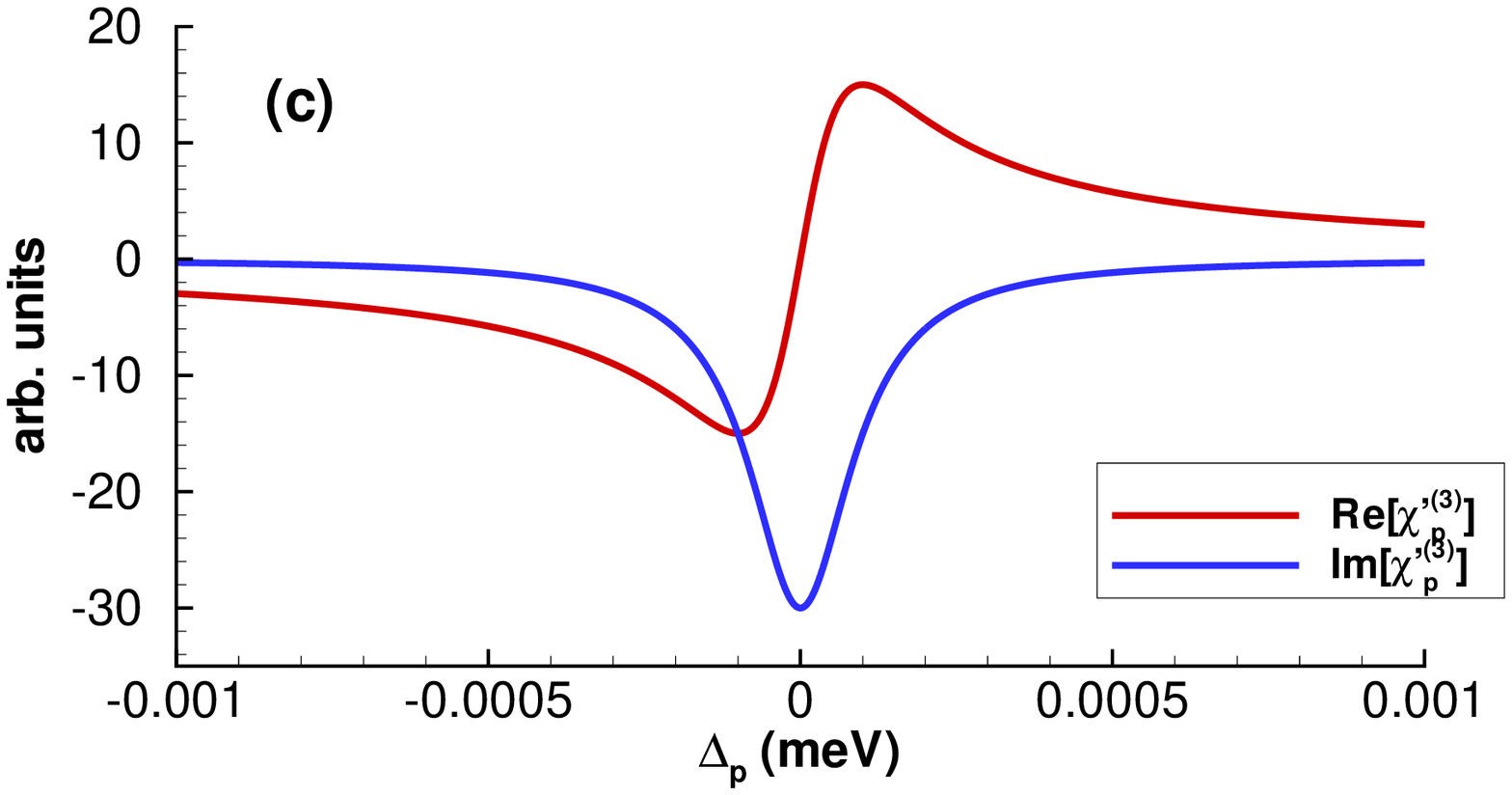}
\includegraphics[width=7cm]{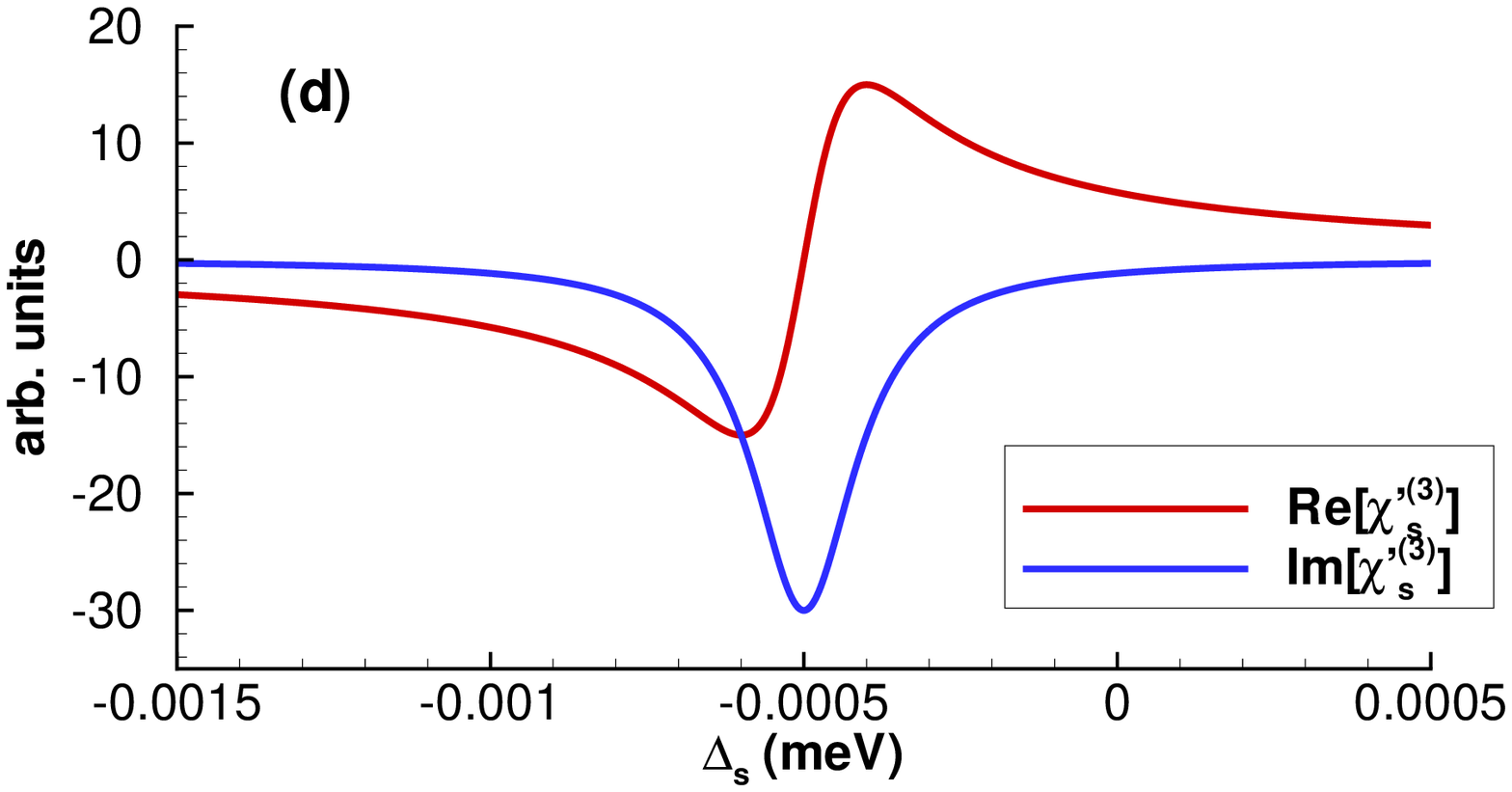}
\caption{(color online) (a) Re[$\chi_{p}^{'(3)}$] and
Im[$\chi_{p}^{'(3)}$] versus $\Delta_{p}$ with (solid curves) and
without (dashed curves) the control field; (b) Re[$\chi_{s}^{'(3)}$]
and Im[$\chi_{s}^{'(3)}$] versus $\Delta_{s}$ with (solid curves)
and without (dashed curves) the control field; (c)
Re[$\chi_{p}^{'(3)}$] and Im[$\chi_{p}^{'(3)}$] versus $\Delta_{p}$
without tunneling by setting $m=q=k=0$; (d) Re[$\chi_{s}^{'(3)}$]
and Im[$\chi_{s}^{'(3)}$] versus $\Delta_{s}$ without tunneling. The
parameters are the same with those in
Fig.~\ref{fig-linear-properties}.}\label{fig-XPM}
\end{figure*}

The explicit forms of the probe and signal third order
susceptibilities associated with cross-Kerr nonlinearity are given
by
\begin{eqnarray}
&&\chi_{p}^{(3,\rm XPM)}=\frac{N|\mu_{13}|^{2}|\mu_{23}|^{2}}{4\hbar^{3}\epsilon_{0}}\chi_{p}^{'(3)},\label{chi3-probe}\\
&&\chi_{s}^{(3,\rm
XPM)}=\frac{N|\mu_{13}|^{2}|\mu_{23}|^{2}}{4\hbar^{3}\epsilon_{0}}\chi_{s}^{'(3)},\label{chi3-signal}
\end{eqnarray}
where $\chi_{p}^{'(3)}$ and $\chi_{s}^{'(3)}$ can be simplified as
\begin{eqnarray}
&&\chi_{p}^{'(3)}=-\frac{T_{p1}}{\cal Z}[T_{p2}+(T_{p3}-T_{p4})\Omega_{c}^{2}+T_{p5}\Omega_{c}^{4}],\label{chip3'}\\
&&\chi_{s}^{'(3)}=-\frac{T_{s1}}{\cal
Z^{*}}[T_{s2}+(T_{s3}-T_{s4})\Omega_{c}^{2}+T_{s5}\Omega_{c}^{4}],\label{chis3'}
\end{eqnarray}
with
\begin{eqnarray*}
&&T_{p1}=d_{51}(d_{41}+mqd_{31})+(m-k)(k-q)\Omega_{c}^{2},\\
&&T_{p2}=d_{51}d_{25}[\sigma_{11}d_{23}d_{24}(d_{41}+mqd_{31})\\
&&\hspace{1cm}-\sigma_{22}d_{31}d_{41}(d_{24}+mqd_{23})],\\
&&T_{p3}=d_{25}[\sigma_{22}(d_{41}+k^{2}d_{31})(d_{24}+mqd_{23})\\
&&\hspace{1cm}+(m-k)(k-q)\sigma_{11}d_{23}d_{24}],\\
&&T_{p4}=d_{51}[\sigma_{11}(d_{24}+k^{2}d_{23})(d_{41}+mqd_{31})\\
&&\hspace{1cm}+(m-k)(k-q)\sigma_{22}d_{31}d_{41}],\\
&&T_{p5}=(k-m)(k-q)[\sigma_{11}(d_{24}+k^{2}d_{23})\\
&&\hspace{1cm}-\sigma_{22}(d_{41}+k^{2}d_{31})],\\
&&T_{s1}=d_{52}(d_{42}+mqd_{32})+(m-k)(k-q)\Omega_{c}^{2},\\
&&T_{s2}=d_{52}d_{15}[\sigma_{11}d_{32}d_{42}(d_{14}+mqd_{13})\\
&&\hspace{1cm}-\sigma_{22}d_{13}d_{14}(d_{42}+mqd_{32})],\\
&&T_{s3}=d_{15}[\sigma_{11}(d_{42}+k^{2}d_{32})(d_{14}+mqd_{13})\\
&&\hspace{1cm}+(m-k)(k-q)\sigma_{22}d_{13}d_{14}],\\
&&T_{s4}=d_{52}[\sigma_{22}(d_{14}+k^{2}d_{13})(d_{42}+mqd_{32})\\
&&\hspace{1cm}+(m-k)(k-q)\sigma_{11}d_{32}d_{42}],
\end{eqnarray*}
\begin{eqnarray*}
&&T_{s5}=(k-m)(k-q)[\sigma_{11}(d_{42}+k^{2}d_{32})\\
&&\hspace{1cm}-\sigma_{22}(d_{14}+k^{2}d_{13})],\\
&&{\cal
Z}=d_{21}[d_{23}d_{24}d_{25}-(d_{24}+k^{2}d_{23})\Omega_{c}^{2}]\\
&&\hspace{1cm}\times[d_{31}d_{41}d_{51}-(d_{41}+k^{2}d_{31})\Omega_{c}^{2}]^{2}.
\end{eqnarray*}

The role of resonant tunneling can be seen from the expressions of
$T_{\alpha\beta}$ ($\alpha=p,s$, $\beta=1,3,4,5$). In the QW
structure under consideration, the symmetric and asymmetric wave
functions of subbands $|3\rangle$ and $|4\rangle$ lead to $m\neq
q\neq k$, which indicates that the resonant tunneling can modify the
optical nonlinearity such as cross-Kerr effect. In
Figs.~\ref{fig-XPM}(a) and \ref{fig-XPM}(b), we illustrate the
evolutions of Re$[\chi_{p,s}^{'(3)}]$ (represent the cross-Kerr
nonlinearities between the probe and signal fields) and
Im$[\chi_{p,s}^{'(3)}]$ (account for the nonlinear absorptions) as
functions of their corresponding detunings with (solid curves) and
without (dashed curves) the control field. All parameters are the
same with those in Figs.~\ref{fig-linear-properties}. Within the
transparency window, both the strengths of cross-Kerr nonlinearities
and nonlinear absorptions of the probe and signal pulses are
enhanced dramatically. Fortunately, the probe and signal nonlinear
absorption peaks are very sharp, and the real parts of the two
cross-Kerr susceptibilities decay much more slowly than their
corresponding nonlinear absorptions. We also notice that in the
present QW structure, $\gamma_{2}$ is dominantly determined by the
electron dephasing rate, smaller $\gamma_{2}$ can be attained by
decreasing the temperature. Hence, the cross-Kerr nonlinearities can
be enhanced much more than that in Ref.~\cite{sun-ol-2007}. More
importantly, positions of the nonlinear absorption peaks can be
controlled by adjusting the detunings $\Delta_{p}$ and $\Delta_{s}$,
which can be seen from Eqs.~(\ref{chip3'}) and (\ref{chis3'}). For
certain detunings, for example $\Delta_{p}=2.9995$~meV and
$\Delta_{s}=3.0$~meV, we have
Re$[\chi_{p}^{'(3)}]\simeq-419.02$~meV$^{-3}$,
Re$[\chi_{s}^{'(3)}]\simeq-417.12$~meV$^{-3}$, and the two negative
cross-Kerr nonlinearities are of the same order of magnitudes.
Therefore, giant cross-Kerr nonlinearities are realized, while the
nonlinear absorptions can be neglected
(Im$[\chi_{p}^{'(3)}]\simeq-10.1$~meV$^{-3}$ and
Im$[\chi_{s}^{'(3)}]\simeq8.7$~meV$^{-3}$). As shown in
Fig.~\ref{fig-XPM}(a) and \ref{fig-XPM}(b), with this set of
parameters, the strengthes of cross-Kerr nonlinearities can be
enhanced with the presence of the control field.

Figures~~\ref{fig-XPM}(c) and \ref{fig-XPM}(d) illustrate the
evolutions of the real and imaginary parts of $\chi_{p}^{'(3)}$ and
$\chi_{s}^{'(3)}$ versus their corresponding detunings with
$m=q=k=0$. In this case, the subband $|4\rangle$ is decoupled, and
the system can hence be described as an inverted-Y-type
configuration. In order to see the effect of resonant tunneling more
clearly, we choose $\Delta_{c}=0$ and $\Omega_{c}\approx2.65$~meV
(similar transparency windows as those with tunneling). The other
parameters are the same with those in Fig.~\ref{fig-XPM}(a) and
\ref{fig-XPM}(b). Within the transparency windows, the enhancement
of cross-Kerr nonlinearities can still be achieved with vanishing
absorptions. However, under the same conditions, the two strengthes
of cross-Kerr nonlinearity are both much less than those with
resonant tunneling. With $\Delta_{p}=0.5$~$\mu$eV and
$\Delta_{s}=0$~meV, we have
Re$[\chi_{p}^{'(3)}]\simeq-5.77$~meV$^{-3}$,
Re$[\chi_{s}^{'(3)}]\simeq5.77$~meV$^{-3}$. Furthermore, the two
cross-Kerr nonlinearities (one positive and one negative) exhibit
destructive effect on the conditional phase shift (details will be
shown later).

A significant interaction is a very essential requirement for
implementation of controlled phase gate between two optical qubits,
where the quantum information is stored in the orthogonal
polarization degree of freedom. In the QW structure, by virtue of
resonant tunneling, such strong interaction can be realized by the
giant cross-Kerr effect. A two-qubit quantum phase gate operation
can be expressed as
$|i\rangle_{p}|j\rangle_{s}\to\exp(i\phi_{ij})|i\rangle_{p}|j\rangle_{s}$,
where $i,j=H,V$ denote the logic qubit basis. The
photon-polarization qubit could be realized easily in the QW
structure considered, where one photon acquires a phase shift
conditioned on the state of another photon. Using polarizing beam
splitters (PBS), we assume the interacting scheme shown in
Fig.~\ref{fig-band-structure} is implemented when the probe and
signal fields have $|H\rangle$
polarization~\cite{petrosyan-job-2005}. After passing through a PBS,
the vertically polarized component of each photon is transmitted,
while the horizontally polarized component is directed into the QW
structure, wherein the two-photon state $|H\rangle_{p}|H\rangle_{s}$
acquires the conditional phase shift. At the output, each photon is
recombined with its vertically polarized component on another PBS.
We assume the probe and signal polarized single-photon wave packets
can be expressed as
\begin{equation}
|\psi_{p,s}\rangle=(|H\rangle_{p,s}+|V\rangle_{p,s})/\sqrt{2},
\end{equation}
which can be written as $|H,V\rangle_{p,s}=\int
d\omega\xi_{p,s}(\omega)a_{H,V}^{\dagger}(\omega)|0\rangle$ with
$\xi_{p,s}(\omega)$ being Gaussian frequency distribution of
incident wave packets centered on $\omega_{p,s}$. Then the
polarization phase gate truth table goes as
\begin{eqnarray}
&&|V\rangle_{p}|V\rangle_{s}=e^{-i(\phi_{p}^{0}+\phi_{s}^{0})}|V\rangle_{p}|V\rangle_{s},\label{eq-truth-table-vv}\\
&&|H\rangle_{p}|V\rangle_{s}=e^{-i(\phi_{p}^{l}+\phi_{s}^{0})}|H\rangle_{p}|V\rangle_{s},\label{eq-truth-table-hv}\\
&&|V\rangle_{p}|H\rangle_{s}=e^{-i(\phi_{p}^{0}+\phi_{s}^{l})}|V\rangle_{p}|H\rangle_{s},\label{eq-truth-table-vh}\\
&&|H\rangle_{p}|H\rangle_{s}=e^{-i(\phi_{p}^{l}+\phi_{s}^{l}+\phi_{p}^{n}+\phi_{s}^{n})}
|H\rangle_{p}|H\rangle_{s},\label{eq-truth-table-hh}
\end{eqnarray}
in which $\phi_{p,s}^{0}=k_{p,s}l$ denotes the trivial vacuum phase
shift with $l$ being the length of the QW structure,
$\phi_{p,s}^{l}=2\pi k_{p,s}l{\rm Re}[\chi_{p,s}^{(1)}]$ linear
phase shift, and $\phi_{p,s}^{n}$ the probe and signal nonlinear
phase shift induced by cross-Kerr effect. The conditional phase
shift can be written as $\phi_{t}^{n}=\phi_{p}^{n}+\phi_{s}^{n}$,
which is only attributed by the cross-Kerr nonlinearity. Provided
the conditional phase shift is not zero, the above set of equations
support a universal quantum phase gate~\cite{ottaviani-prl-2003}.
For Gaussian probe and signal pulses of time durations $\tau_{p,s}$,
and with peak Rabi frequencies $\Omega_{p,s}^{0}$, solving the
propagation equations gives the nonlinear cross-phase shift
$\phi_{p,s}^{n}$\cite{ottaviania-ejpd-2006}
\begin{eqnarray}
&&\phi_{p}^{n}=\frac{2\omega_{p}l}{c}\frac{\hbar^{2}|\Omega_{s}^{0}|^{2}}{|\mu_{23}|^{2}}\frac{{\rm
erf}(\zeta_{p})}{\zeta_{p}}{\rm Re}[\chi_{p}^{(3)}],\\
&&\phi_{s}^{n}=\frac{2\omega_{s}l}{c}\frac{\hbar^{2}|\Omega_{p}^{0}|^{2}}{|\mu_{13}|^{2}}\frac{{\rm
erf}(\zeta_{s})}{\zeta_{s}}{\rm Re}[\chi_{s}^{(3)}],
\end{eqnarray}
\begin{figure}
\includegraphics[width=8cm]{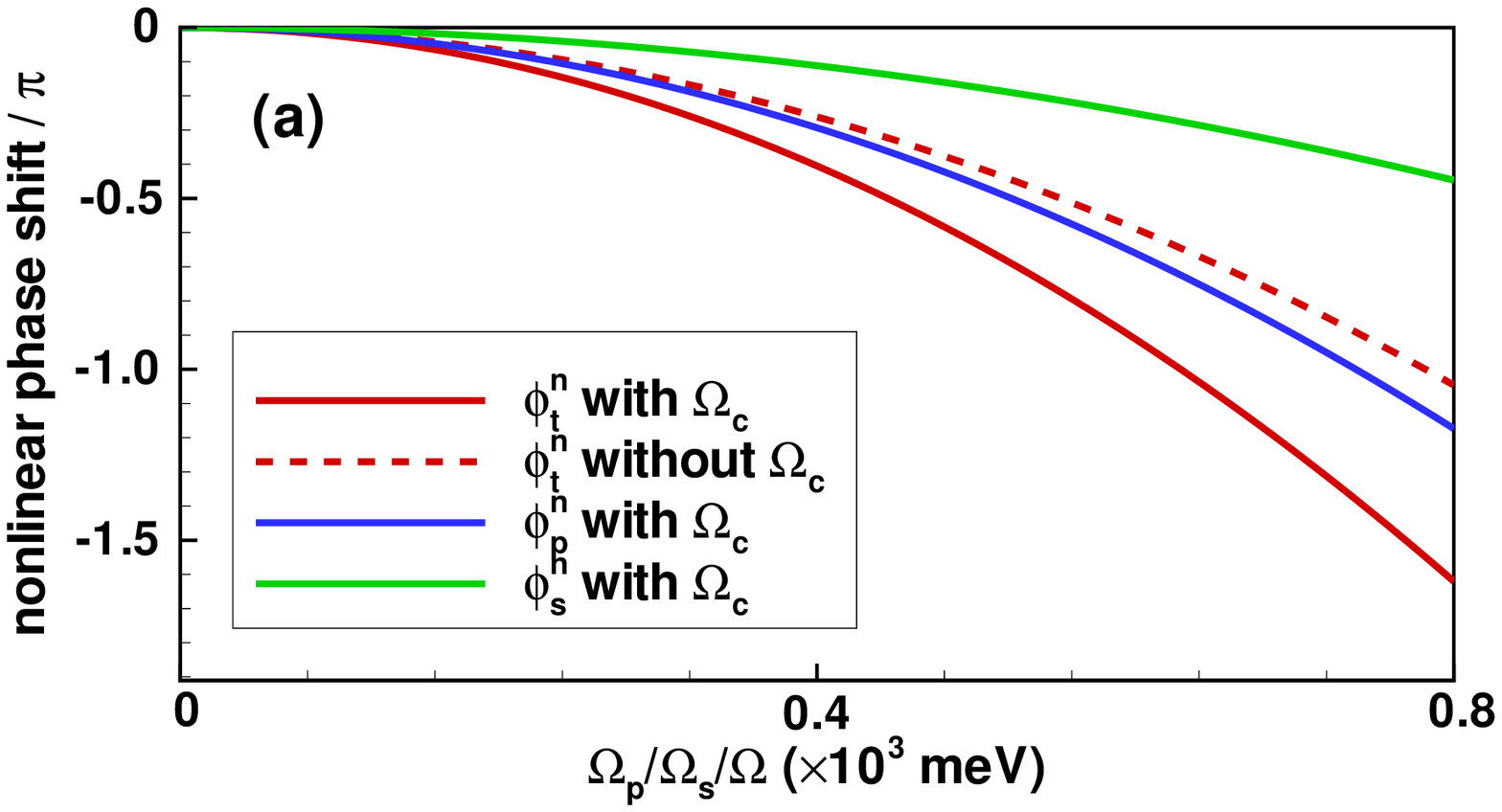}
\includegraphics[width=8cm]{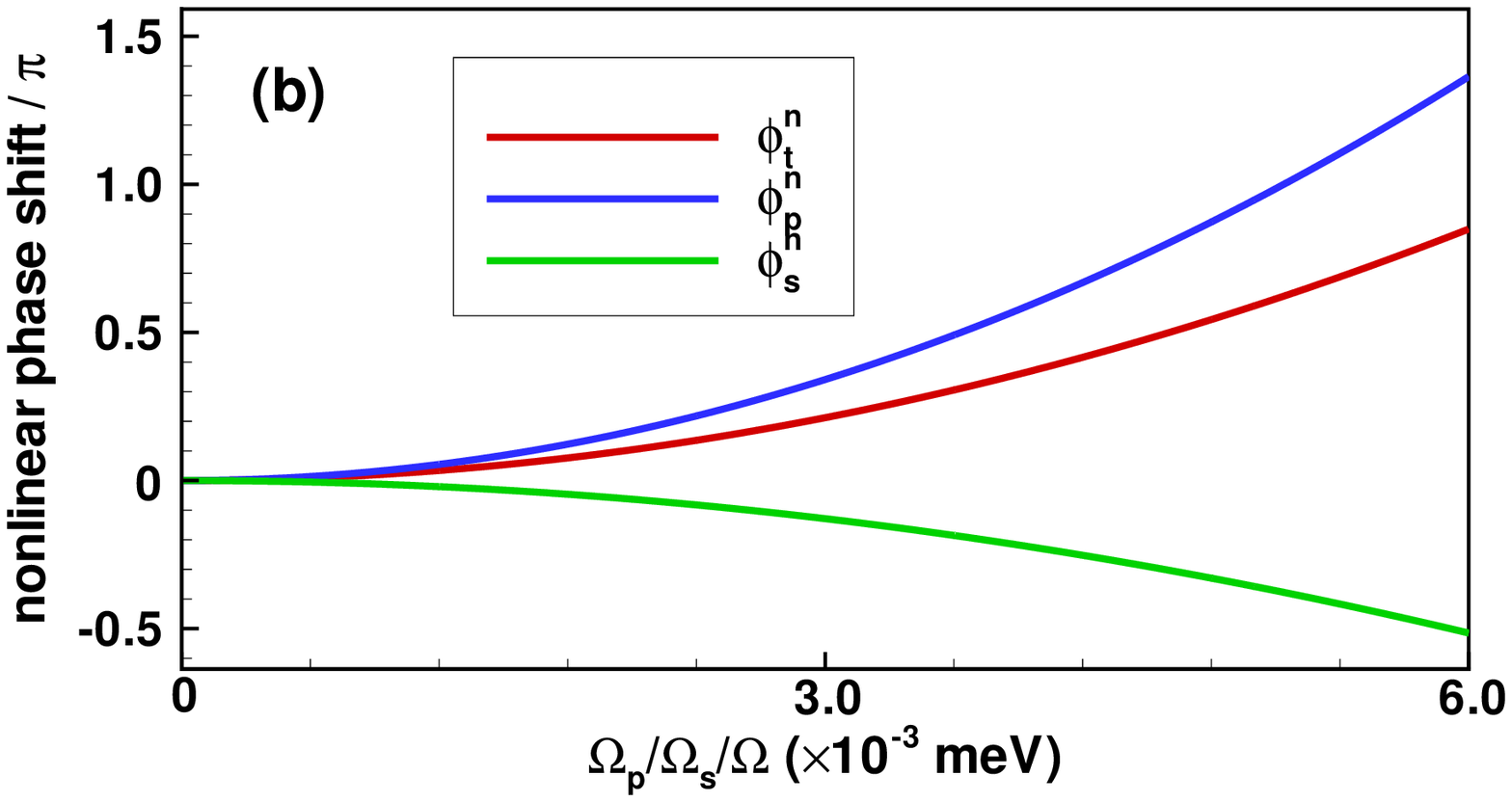}
\caption{(color online) (a) $\phi_{p}^{n}$, $\phi_{s}^{n}$, and
$\phi_{t}^{n}$ with and without the control field versus the Rabi
frequency $\Omega_{p}=\Omega_{s}=\Omega$; (b) $\phi_{p}^{n}$,
$\phi_{s}^{n}$, and $\phi_{t}^{n}$ versus the Rabi frequency
$\Omega_{p}=\Omega_{s}=\Omega$ with $m=q=k=0$. The length of QWs is
taken as $1.0$~mm, and the other parameters are same as those in
Fig.~\ref{fig-linear-properties}.}\label{fig-phase-shift}
\end{figure}
where
$\zeta_{p}=[(1-v_{g}^{p}/v_{g}^{s})\sqrt{2}l]/(v_{g}^{p}\tau_{s})$,
and $\zeta_{s}$ can be obtained from $\zeta_{p}$ upon interchanging
$p\leftrightarrow s$. erf$(\zeta)$ represents the error function.
The nonlinear phase shift acquired by the probe and signal pulses
propagating through the QW structure can be controlled by the signal
and probe pulses intensity. In Figs.~\ref{fig-phase-shift} (a) and
(b), we plot the evolutions of the probe nonlinear phase shift
$\phi_{p}^{n}$, the signal nonlinear phase shift $\phi_{s}^{n}$, and
the conditional phase shift $\phi_{t}^{n}$ as functions of
$\Omega_{p}^{0}=\Omega_{s}^{0}=\Omega$ with (a) and without (b)
resonant tunneling. The length of QW structure is taken as
$l=1.0$~mm. With resonant tunneling, ${\rm Re}[\chi_{p}^{(3,\rm
XPM)}]\cdot{\rm Re}[\chi_{s}^{(3,\rm XPM)}]>0$ leads to
$\phi_{p}^{n}\cdot\phi_{s}^{n}>0$, which indicates the constructive
effect of the probe and signal nonlinear phase shift on the
conditional phase shift (see Fig.~\ref{fig-phase-shift}(a)).
$\phi_{t}^{n}=\pi$ can be achieved with
$\Omega_{p}=\Omega_{s}=\Omega\approx6.22\times10^{-4}$~meV. The
probe and signal pulses can have a mean amplitude of about one
photon when these beams are focused or propagate in a tightly
confined waveguide. With these parameters, the corresponding
intensities of the probe and the signal pulses are, respectively,
given by $I_{p}\approx4.35$~mW~cm$^{-2}$ and
$I_{s}\approx6.94$~mW~cm$^{-2}$. We remark that the intensities of a
single probe and signal photons per 0.1~ns on the area of
$1$~$\mu$m$^{2}$ are $I_{p}\approx27.3$~mW~cm$^{-2}$ and
$I_{s}\approx22.5$~mW~cm$^{-2}$, respectively. The numerical
findings indicate that our semiconductor QW structure can indeed
make a polarization photonic controlled phase gate with a
$\pi$-conditional phase shift possible with single-photon wave
packets. In Fig.~\ref{fig-phase-shift}(a), we also illustrate the
positive effect of the control field on the conditional phase shift.
Without resonant tunneling, ${\rm Re}[\chi_{p}^{(3,\rm
XPM)}]\cdot{\rm Re}[\chi_{s}^{(3,\rm XPM)}]<0$ exhibits the
destructive effect on conditional phase shift (see
Fig.~\ref{fig-phase-shift}(b)). In this case, the conditional phase
shift on order of $\pi$ can be obtained with more than one photons,
i.e., $\Omega_{p}=\Omega_{s}=\Omega\approx6.5\times10^{-3}$~meV
($I_{p}\approx0.48$~W~cm$^{-2}$ and $I_{s}\approx0.76$~W~cm$^{-2}$).

\begin{figure}
\includegraphics[width=8cm]{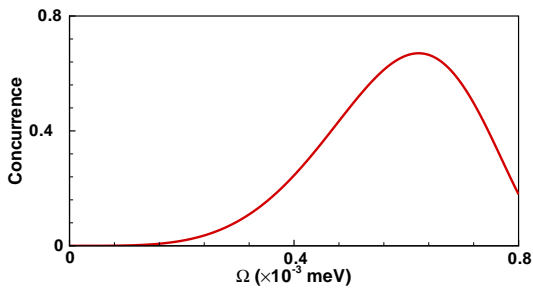}
\caption{(color online) The concurrence versus the Rabi frequency
$\Omega_{p}=\Omega_{s}=\Omega$. The other parameters are same as
those in Fig.~\ref{fig-linear-properties}.}\label{fig-concurrence}
\end{figure}

We now turn to the problem of entangled-photon states in our QW
structure by using the entanglement of formation. Starting from the
truth table
(Eqs.~(\ref{eq-truth-table-vv})-(\ref{eq-truth-table-hh})), the
degree of entanglement of two polarized photons state can be
computed. For an arbitrary two-qubit system, the degree of
entanglement is defined by~\cite{wootters-prl-1998}
\begin{equation}
E_{F}(C)=h\left(\frac{1+\sqrt{1-C^{2}}}{2}\right),\label{entanglement}
\end{equation}
where $h(x)=-x\log_{2}(x)-(1-x)\log_{2}(1-x)$ is Shannon's entropy
function, and concurrence $C$ is given by
\begin{equation}
C(\hat{\rho})={\rm
max}\{0,\lambda_{1}-\lambda_{2}-\lambda_{3}-\lambda_{4}\},\label{concurrence}
\end{equation}
with $\lambda_{i}$ ($i=1-4$) being the square roots of the
eigenvalues of
$\hat{\rho}(\hat{\sigma}_{y}^{p}\otimes\hat{\sigma}_{y}^{s})
\hat{\rho}^{*}(\hat{\sigma}_{y}^{p}\otimes\hat{\sigma}_{y}^{s})$ in
descending order. Here $\hat{\rho}^{*}$ denotes the complex
conjugation of the output state density matrix $\hat{\rho}$, and
$\hat{\sigma}_{y}^{j}$ ($j=p,s$) is the $y$-component of the Pauli
matrix. The concurrence $C$ can be taken as a kind of measure of
entanglement since $E_{F}(C)$ is a monotonic increasing function of
$C$. As an example, we plot the evolution of the concurrence versus
the Rabi frequency $\Omega_{p}=\Omega_{s}=\Omega$ in
Fig.~\ref{fig-concurrence}. With the set of parameters in
Fig.~\ref{fig-phase-shift}(a), the maximum degree of entanglement
obtained in our proposal can be as large as $E_{F}\approx0.55$ (in
Ref.~\cite{yang-oe-2008}, $E_{F}\approx0.35$) with
$\Omega\approx6.22\times10^{-4}$~meV, corresponding to the point of
the conditional phase shift $\phi_{t}^{n}\approx\pi$.

\section{conclusions}

In the literature several semiconductor QW structures have been
suggested to investigate the possibility of quantum phase
gate~~\cite{yang-oe-2008,hao-josab-2010}. We should point out that,
in asymmetrical N-type~\cite{yang-oe-2008} or
ladder~\cite{hao-josab-2010} configurations, the probe pulse
propagates with slow light because of EIT, while the signal pulse
possesses a nonzero Kerr nonlinearity only. The group velocity
matching can only be satisfied by controlling the signal detuning
when the signal field is continuous wave. However, this is not
desirable for photonic controlled phase gate. This is a consequence
of the asymmetry configurations. In the present study, within the
transparency window considered, the influence of group velocity
dispersion can be ignored safely. That is to say, it is possible
that the probe and signal wave packets propagate in QW structure
with group velocity matching and higher stability.

In conclusion, we have designed a double QW structure to achieve
strongly interacting and highly entangled photons. This structure
combines the resonant tunneling with the advantages of inverted-Y
type scheme. By virtue of resonant tunneling, not only the strength
of cross-Kerr nonlinearities can be enhanced dramatically with
vanishing linear and nonlinear absorptions simultaneously, but also
the effect of cross-Kerr nonlinearities of the probe and signal
pulses on the conditional phase shift can be changed from
destructive to constructive. Our numerical findings confirm that it
is possible to achieve nonlinear phase shift on order of $\pi$ at a
single photon level. Based on such important features, we have
demonstrated that it is possible to produce highly entangled photon
pairs and construct polarization qubit quantum phase gates. For the
probe and the signal pulses, the interacting scheme is symmetric,
and thus yields equal group velocities by adjusting the initial
distribution of electron. We believe that the present study may be
useful for guiding experimental realization of electroptically
modulated devices and facilitating more practical applications in
solid quantum information processing.

\begin{acknowledgements}
We thank the financial support form the Fundamental Research Funds
for the Central University under Grant No. GK201003003 and the Open
Fund from the SKLPS of ECNU, as well as the National Research
Foundation and Ministry of Education, Singapore under academic
research grant No. WBS: R-710-000-008-271. The author (G.X.H) would
like to acknowledge the support of the NSF-China under Grant No.
10874043, and the author (S.Q.G) acknowledges funds from the
National Natural Science Foundation of China under Grant No.
60978013, as well as the support of Shanghai Commission of Science
and Technology with Grant No. 10530704800.
\end{acknowledgements}

\end{document}